\documentclass[reprint,amsmath,amssymb, aps]{revtex4-2}
\usepackage[english]{babel}
\usepackage[utf8]{inputenc}
\usepackage{adjustbox}
\usepackage{afterpage}
\usepackage{siunitx}
\usepackage{float}
\usepackage{xcolor}

\newcommand{\emphR}[1]{(\emph{#1})}

\begin{document}
\title{Droplet-on-demand using a positive pressure pulse}

\author{Mathieu Oléron$^{1,2}$}
\author{Grégoire Clement$^{1,2}$}
\author{Samuel Hidalgo-Caballero$^{1,2}$}
\author{Masoodah Gunny$^{1,2}$}
\author{Finn Box$^{1,2,3}$}
\author{Matthieu Labousse$^{1}$}
\email{matthieu.labousse@espci.fr}

\author{Joshua D. McGraw$^{1,2}$}
\email{joshua.mcgraw@espci.fr}

\affiliation{$^{1}$Gulliver UMR 7083 CNRS, ESPCI--PSL, 10 rue Vauquelin, 75005 Paris, France}
\affiliation{$^{2}$IPGG, 6 rue Jean-Calvin, 75005 Paris, France}
\affiliation{$^{3}$Department of Physics \& Astronomy, University of Manchester, Manchester M13 9PL, United Kingdom}

\date{\today} 


\begin{abstract}
Droplet generation under steady conditions is a common microfluidic method for producing biphasic systems. However, this process works only over a limited range of imposed pressure: beyond a critical value, a stable liquid jet can instead form. Furthermore, for a given geometry the pressure conditions set both the generation rate of droplets and their volume. Here, we report on-demand droplet production using a positive pressure pulse to the dispersed-phase inlet of a flow-focusing geometry. This strategy enables confined droplet generation within and beyond the pressure range observed under steady conditions, and decouples volume and production rate. In particular, elongated plugs not possible under steady conditions may be formed when the maximal pressure during the pulse reaches the jet regime. The measured volume of droplets-on-demand, as well as the onset of droplet generation are both captured with a simple model that considers hydraulic resistances. This work provides a strategy and design rules for processes that require individual droplets or elongated plugs in a simple microfluidic chip design. 
\end{abstract}


\maketitle


\section{Introduction}

 Droplet-based microfluidics finds application in many fields, from materials engineering to medical applications~\cite{seemann2011droplet, zhu2017passive}. For example, producing droplets is the first step to generate microcapsules~\cite{datta201425th, lee2016microfluidic}, microfibers~\cite{nunes2013dripping} and micro- or nano-particles~\cite{kong2012droplet, song2008microfluidic, jahn2008preparation, martz2012microfluidic}. These components then find applications in the food~\cite{schroen2021droplet} and cosmetics industries~\cite{park2021microfluidic}.
Droplet-based microfluidics also enables biological assays~\cite{theberge2010microdroplets, vyawahare2010miniaturization} for immobilization of reagents or compartmentalization. 

\begin{figure}[t!]
\centering
\includegraphics[scale=0.991]{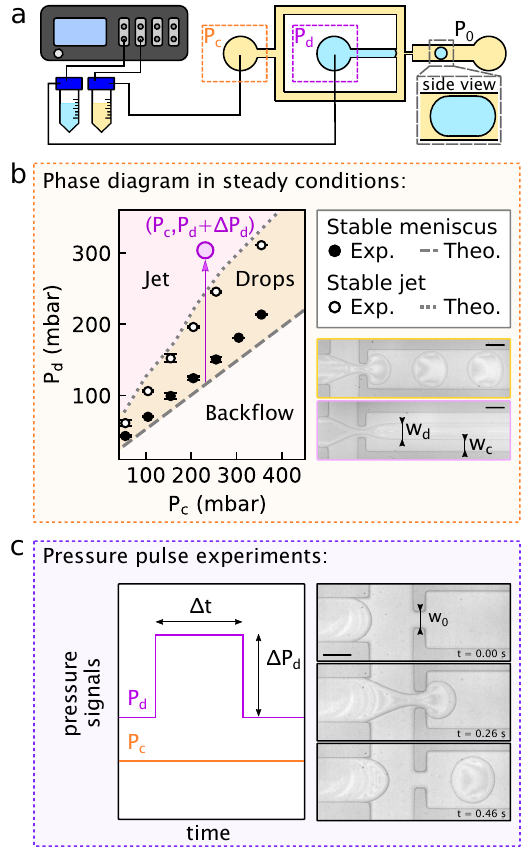} 
\caption{\textbf{Experimental setup, steady state and pulsed droplet generation}. (a) Setup: a pressure controller injects two fluids into a flow-focusing microfluidic chip. $P_{\rm c}$ and $P_{\rm d}$ are the pressures in the continuous phase and dispersed phase channels, respectively; the outlet is at atmospheric pressure $P_0$. (b) Steady-condition phase diagram. Orange patch and picture: droplet production regime $(P_{\rm d,0}~\text{(grey dashed line)} < P_{\rm d} < P_{\rm d,jet}~\text{(grey dotted line)}$). Pink patch and picture: stable jet co-flow ($P_{\rm d} > P_{\rm d,jet}$). The respective widths of the continuous and dispersed flow are $w_{\rm c}$ and $w_{\rm d}$. White patch ($P_{\rm d} < P_{\rm d,0}$): backflow region (c) Pressure signals during a pulse experiment with a square signal of amplitude $\Delta P_\mathrm{d}$, as indicated by the purple arrow in (b), and duration $\Delta t$. At right are shown micrographs of the result of the pressure pulse with pressures $\mathbf{P} = \{155,98,100\}$~\si{\milli\bar} and $\Delta t = 0.10$~\si{\second}. The neck of the cross-junction has a width $w_{\rm 0} = 50$~\si{\micro\meter}. Scale bars: $100$~\si{\micro\meter}.}
\label{fig:setup}
\end{figure}

To create droplets, two fluids in separate branches of a microfluidic chip are normally made to meet at a junction, and usually under constant imposed pressure or flow rate. In a flow-focusing configuration as shown in Fig.~\ref{fig:setup}(a), droplet production requires appropriate pressures to be applied at the inlets for the device to function as a droplet generator. When the pressure in the dispersed phase channel (\emph{i.e.}, the phase of the droplets) exceeds the pressure at which the interface between the two fluids remains stable, determined by fluidic resistance of the channels and the Laplace pressure~\cite{BruusTheoreticalMicrofluidics2007}, the dispersed phase flows towards the exit channel. If the pressure is below a second threshold that we call the jetting pressure, the liquid/liquid interface ruptures to create a droplet. This latter droplet is carried away by the continuous phase, and the process repeats under the imposed constant inlet pressures. If the pressure is above the jetting one, an stable stream of the dispersed phase flows together with the continuous phase towards the exit. The jet remains stable when confined in one direction~\cite{guillot2008stability}. 

The droplet-generation method in between two pressure thresholds described in the previous paragraph may lead to high throughput, from \emph{e.g.} $0.1$~\si{\hertz} to $10$~\si{\kilo\hertz}~\cite{garstecki2004formation, takeuchi2005axisymmetric, yobas2006high, tirandazi2017liquid, lashkaripour2019performance}. However, one drawback of working at constant imposed pressure is that it is difficult to control independently droplet size and the frequency at which droplets are created. Indeed, for some applications it may be desirable to attain a full channel containing only a single droplet of given size, rather than a continuous stream of droplets. Examples where an empty channel may be necessary include the study of dissolving~\cite{Sun2011, Cubaud2021} or active droplets~\cite{DeBlois2021} for which flushing the exit channel between droplet passages is commonly employed, or cases for which non-periodic or conditional droplet production is necessary (\emph{e.g.}, \emph{in-situ} screening with droplet generation after a positive signal). Furthermore, one solution to effectively increase droplet separation under steady conditions is to impose additional flow from extra side channels~\cite{Reichert2019}, yet this necessarily accelerates the droplet under study, and may not lead to a droplet-free exit channel. Besides, the presence of a train of droplets can significantly modify a channel's hydraulic resistance~\cite{Vanapalli2009}, while negligible changes may arise in the case of one single droplet. 

To satisfy these constraints, producing droplets on-demand indeed makes it possible to decouple size from generation frequency, with the latter reaching arbitrarily low values. To this end, different groups have used electric~\cite{he2005electro, he2006effects, zhang2008manipulations, niu2009generation, gu2011microfluidic} or magnetic fields~\cite{kahkeshani2016drop}, a laser pulse~\cite{Wu2012PulsedLaser}, surface acoustic waves~\cite{collins2013surface,brenker2016chip,brenker2020demand}, piezoelectric actuators~\cite{xu2008drop,bransky2009microfluidic,shemesh2011coalescence}, and on-chip~\cite{galas2009active, lin2008demand, lee2009predictive, zeng2009microvalve, churski2010droplet, guo2010valve} or off-chip microvalves~\cite{moon2015microfluidic, zhou2014facile, churski2010high,dolega2012iterative, yu2015demand, jakiela2014generation, jung2010demand}. While effective, these methods require additional fields or specialized equipment. 

One of the simplest methods for producing individual droplets is to use a pressure pulse on either the dispersed or continuous phases. Nevertheless, to the best of our knowledge, only a few groups have applied this strategy to generate droplets one by one~\cite{hamidovic2020off, teo2017negative, fink2020design, totlani2020scalable}. These studies demonstrated that the pulsed-pressure method is straightforward to implement using a pressure controller. The pulse strategy provides a convenient means to program trains of droplets~\cite{hamidovic2020off, fink2020design} or control the mixing of two fluids~\cite{hamidovic2020off, totlani2020scalable}. The use of a chamber and dual-pulse droplet generation provides better control over the volume injected~\cite{totlani2020scalable}. However, a rationalization of the shapes observed as a function of the control parameters and pulse features is still lacking. 
 
In this paper, we explore the conditions at which a single droplet results from a  pressure pulse.  We applied pulses with various amplitudes and durations at different initial pressures to the dispersed phase inlet of a flow-focusing chip. Our observations show that the droplet-on-demand (DoD) morphologies are closely linked to the droplet production and stable jet regimes observed in steady conditions. Finally, a simple model based on the hydraulic resistances of the different parts of the flow-focusing geometry ---similar to Teo \emph{et al.}'s analysis~\cite{teo2017negative} describing experiments wherein a negative pressure pulse on the continuous phase was used--- is able to predict \emphR{i} the minimal pulse time necessary to produce a droplet, and \emphR{ii} the corresponding volume. 

The present article is organised as follows: in the next section, we describe the methods and materials used to apply pressure pulses to stable meniscii in a flow-focusing geometry. Then, the results and discussion section in turn presents: an overview of the steady-state phase diagram for which pressures are continuously applied; the dependence of droplet morphology on the initial pressures, the pulse amplitude and its duration; and finally, a simple scaling analysis leading to design rules for minimal droplet generation time and its corresponding volume.


\section{Methods}

\subsection{Microfluidic chip: fabricaton, materials and characteristics} 
We prepared microfluidic chips with poly-dimethylsiloxane (PDMS, Momentive RTV 615) using typical soft lithography techniques~\cite{xia1998soft}, pouring the uncrosslinked elastomer onto a textured and hydrophobized (trichloro(1H, 1H, 2H, 2H-perfluorooctyl)silane, Sigma Aldrich) silicon wafer. This latter silanization was used to facilitate peeling the cured PDMS off the wafer. To prepare the bottom surface of the channel, we cleaned the glass slides with laboratory-grade acetone and ultra-pure water (milliQ, 18 M$\Omega$.\si{\cm}), and dried them with compressed air. Then, we spin coated a layer of PDMS, cured it for $4$~\si{\hour} at $70$\ \si{\celsius}. This surface was bound to the microfluidic chip, which was similarly cured, after plasma activation (Femto Science, CUTE). To restore hydrophobicity of the PDMS surfaces, we annealed the chip at $90$\ \si{\celsius} for $15$\ \si{\hour}~\cite{pascual2019wettability}.

The geometry used was a flow-focusing one~\cite{garstecki2004formation, garstecki2005mechanism, teo2017negative, dollet2008role} shown in Fig.~\ref{fig:setup}(a). The three principal sections of the chip are: the dispersed-phase entrance (blue channel, subscript ``d'' in the following), continuous-phase entrance (yellow channel surrounding the former one, subscript ``c'') and the exit (channel schematically containing the drop, subscript ``x''). The channel widths were respectively $w = 150$, $150$ and $200$~\si{\micro\meter}; their height was measured using a mechanical profilometer (Dektak, Veeco) as $h = 28.0\pm0.3$~\si{\micro\meter}. Given the confinement in the out-of-plane dimension, all the droplets observed here have pancake-like morphologies as shown in the side-view panel in Fig.~\ref{fig:setup}(a). The dispersed and continuous phases meet at a cross-junction, with a constriction of width $w_{\rm 0} = 50$~\si{\micro\meter} being crossed before arrival to the exit channel. We note that, while the focus here is on flow focusing geometries with a positive, dispersed-phase pressure pulse, we verified that a droplet could be produced using: (\emph{i}) a positive, dispersed-phase pressure pulse with a T-junction geometry, and (\emph{ii}) a negative, continuous-phase pressure pulse in a flow-focusing geometry.

For all experiments, the continuous phase was a $1$\ \si{\percent}-by-weight mixture of a perfluorinated oil (Perfluoroalkylether, Krytox 1506, purchased from Sigma Aldrich) with a surfactant (Perfuoro polyether carboxylic acid, Krytox 157 FSL); ultra-pure water was used as the dispersed phase. All solutions were passed through a $2$-\si{\micro\meter} filter before use to avoid dust contamination in the chip. The liquid densities were measured to be $\rho_{\rm c} = 1.89 \pm 0.04$~\si{\gram}.\si{\per\milli\liter} and $\rho_{\rm d} = 1.00 \pm 0.04$~\si{\gram}.\si{\per\milli\liter}. The dynamic viscosities determined with a rheometer (MCR 302, Anton Paar) in double-Couette geometry at room temperature, were found to be $\eta_{\rm d} = 1.0 \pm 0.1$~\si{\milli\pascal}.\si{\second} and $\eta_{\rm c} = 88 \pm 3$~\si{\milli\pascal}.\si{\second}. Finally, the surface tension between the two liquid phases was determined using the pendent droplet method~\cite{daerr2016pendent_drop} to be $\gamma = 5\pm3$~\si{\milli\newton}.\si{\per\meter}. 

A particular quantity of interest in the context of droplet production is the hydraulic resistance of a channel branch, $R$. In analogy with the electrical resistance in Ohm's law, we have $R= \Delta P/Q$ with $\Delta P$ the pressure drop along the channel section, playing the role of the voltage, and $Q$ the volumetric flow rate taking the role of the current. This hydraulic resistance $R$ depends on the channel geometry (length $L$, width $w$ and height $h$) and on the fluid viscosity $\eta$.  In the case of a Hagen-Poiseuille flow~\cite{oh2012design} and a rectangular channel~\cite{cornish1928flow, BruusTheoreticalMicrofluidics2007}, the resistance is well-approximated by 

\begin{equation}
R = \dfrac{12 \eta L}{w h^3 \left( 1 -  0.63\dfrac{h}{w} \right) }\ .
\label{eq:HydroRes}
\end{equation}
The estimated hydraulic resistances of the dispersed, continuous-phase-entrance and the exit channels are computed and shown in Tab.~\ref{tab:ChipGeo}.

\begin{table}
\begin{center}
\begin{adjustbox}{width=0.99\columnwidth}
\begin{tabular}{l c c c c c}
\hline
 Section & $h$ & $w$ & $L$ & $\eta$ & $R$  \\
 & \si{\micro\meter} & \si{\micro\meter} & \si{\milli\meter} & \si{\milli\pascal}.\si{\second} & \si{\kilo\pascal }.\si{\second }.\si{\per\nano\liter}\\
\hline
\hline
Dispersed & $28.5(3)$ & $150(3)$ & 9.2(1) & $1.0(1)$ &  $0.041(4)$\\
Continuous& $28.5(3)$ & $150(3)$ & 12.0(1) & $88(3)$ & $2.25(7)$\\
Exit & $28.5(3)$ & $200(3)$ & 8.0(1) & $88(3)$ & $2.08(7)$\\ 
\hline
\end{tabular}
\end{adjustbox}
\end{center}
\caption{\textbf{Geometrical parameters of the chip.} The hydraulic resistances of the continuous and exit sections $R_{\rm c}$ and $R_{\rm x}$ are calculated with the viscosity $\eta_{\rm c}$. The hydraulic resistance of the dispersed section $R_{\rm d}$ is calculated with the viscosity $\eta_{\rm d}$. Errors in the last digit are shown in parentheses for all quantities. }
\label{tab:ChipGeo}
\end{table}

\vspace{3mm}

\subsection{Pressure variation for droplet generation}
A pressure controller (Elveflow$^{\small{\textregistered}}$, OB1 MK4+) was used to set the pressures inside $15$~\si{\milli\liter} centrifuge tubes (Falcon 352097, Corning). These tubes, containing the continuous and dispersed phases, were connected to the inlets of the microfluidic chip as shown in Fig.~\ref{fig:setup}(a). For the steady-pressure case, in a first stage we determined the cutoff pressures for which, respectively, the meniscus was steady and for which a stable jet is formed in the exit channel as shown in Fig.~\ref{fig:setup}(b). 

The second stage of the DoD method consists of setting the pressures so that the water-oil interface remains stable close to the cross-junction. In this configuration, only the continuous phase flows towards the outlet. Then, the pressure applied to the dispersed phase is suddenly increased by an amount $\Delta P_{\rm d}$ for a time $\Delta t$ as schematically indicated by the purple arrow and datum in Fig.~\ref{fig:setup}(b) and in Fig.~\ref{fig:setup}(c). Our pressure controller has a time resolution of approximately $0.1$~s; therefore, a maximal droplet production frequency of order $10\ \si{\hertz}$~\cite{hamidovic2020off} could be expected using this method, which is thus not adapted to applications requiring rates in the kHz frequency range~\cite{Agresti2010, Jeong2012}. The dispersed phase is nevertheless pushed into the cross junction, and squeezed by the flow of the continuous phase. If the pulse is long enough, the dispersed phase eventually breaks to form a droplet as seen in Fig.~\ref{fig:setup}(c), with the droplet carried away by the continuous flow. Before proceeding to the creation of another droplet, a sufficient delay time is allowed for the exit channel to be droplet-free. Hence, modifications of the hydraulic resistance of this channel and thus pressure changes at the junction point are avoided, which would be a factor with one or several moving droplet inside the exit channel~\cite{jousse2005compact, fuerstman2007pressure, schindler2008droplet}. This precaution ensures reproducible initial conditions. 

In our experiments, we varied the parameters of the pressure pulse to explore the resulting droplet morphologies. These former are described using a pressure triplet $\mathbf{P} = \{P_{\rm c},P_{\rm d}, \Delta P_{\rm d}\}$ and the pulse duration $\Delta t$. First, we fixed $\mathbf{P}$ and varied $\Delta t$. Then, $\Delta P_{\rm d}$ was swept for a selection of different $\Delta t$. Finally, the initial pressures $\{P_{\rm c},P_{\rm d}\}$ were changed and the previous protocols were repeated. In our experiments, $\Delta P_{\rm d} \in \lbrace 5, 20, 50, 100, 160 \rbrace$~\si{\milli\bar} and $P_{\rm c}$ is set sequentially to one of $\lbrace 55, 155, 255, 355 \rbrace$~\si{\milli\bar} after correction with the hydrostatic pressures, with $P_{\rm d}$ determined according to the stability criterion described in the next section. The applied pulses $\mathbf{P}$ were chosen to explore the steady phase diagram, both in the droplet and jet production regimes.


\section{Results and discussion}

\subsection{Phase diagram in steady conditions}\label{SectSteady}
In this subsection, we recall the operating regimes for droplet and jet production under steady conditions (\emph{i.e.} for which $\Delta t\rightarrow \infty$). Such discussions appear in several other works, notably the one by Teo \emph{et al.}~\cite{teo2017negative}. Therefore, we discuss only the main ingredients serving as the basis for the interpretation of our pulsed-pressure experiments: the meniscus stability, the droplet production regime, and the transition from droplet production to jetting. 

\begin{figure}[t!]
\centering
\includegraphics[scale=1]{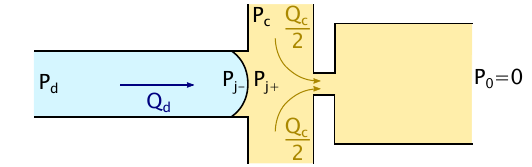} 
\caption{\textbf{Sketch of the cross-junction}. The pressure controller imposes $P_{\rm c}$ to the continuous phase inlet, and $P_{\rm d,max} = P_{\rm d} + \Delta P_{\rm d}$ to the dispersed phase inlet. $Q_{\rm c}$ and $Q_{\rm d}$ are the resulting flow rates. The pressure cost to cross the curved interface is the Laplace pressure $\Delta P_{\rm L} = P_{\rm j^{-}} - P_{\rm j^{+}}$. The outlet is at atmospheric pressure $P_{0}$.}
\label{fig:CrossSketch}
\end{figure}

In order to stabilize the meniscus for a given $P_{\rm c}$, there exists a single pressure $P_{\rm d,0}$ for which the dispersed-phase flow rate $Q_\mathrm{d}=0$. These stability conditions are indicated with filled circles in Fig \ref{fig:setup}(b). In this steady condition, a balance at the junction shown in Fig.~\ref{fig:CrossSketch} is given by the dispersed-phase pressure diminished by the Laplace one, counterbalanced by the continuous-phase pressure at the junction. We denote this Laplace pressure jump as $\Delta P_{\rm L} = P_{\rm j^-} - P_{\rm j^+} = 2\gamma (1/h + 1/w)$, where $P_{\rm j^\mp}$ are the pressures in the dispersed/continuous phases on either side of the meniscus. Equating the pressures from the dispersed and continuous phase inlets to the positive side of the junction, and using mass conservation of the continuous phase between the entrance and exit channels, the stability criterion is
\begin{equation}
P_{\rm d,0} = P_{\rm c} \left( \frac{R_{\rm c}}{R_{\rm x}}+1 \right)^{-1}+ \Delta P_{\rm L}\ .
\label{eq:StabMen}
\end{equation} 
Eq. \ref{eq:StabMen} compares reasonably well with the experimental stable meniscus conditions, even if it leads to a small underestimation for larger $P_{\rm c}$ (dashed line in Fig.~\ref{fig:setup}(b)).
These discrepancies may be due to the deformability of the channel~\cite{guyard2022elastohydrodynamic}, and the associated change in the hydraulic resistances~\cite{teo2017negative}. 

Then, on the one hand, if $P_{\rm d}<P_{\rm d,0}$ there is a back flow of the continuous phase into the dispersed phase channel, as indicated by the lower, white patch in Fig.~\ref{fig:setup}(b). On the other hand, if $P_{\rm d}$ is above $P_{\rm d,0}$, water flows out towards the outlet channel, enabling droplet production as indicated by the orange patch in Fig.~\ref{fig:setup}(b). The forces responsible for the pinch-off of the interface depend on the pressures applied~\cite{zhu2017passive}. In the so-called ``squeezing regime,'' the dispersed phase entirely fills the channel junction and blocks the flow of continuous phase, increasing the pressure at this latter at the junction. The resulting pressure gradient in the continuous flow leads to breakup once it overcomes the pressure of the dispersed phase. As $P_{\rm c}$ is increased, the continuous flow shears the oil/water interface driving a pinch-off event; this scenario is called the ``dripping'' regime. The transition from the squeezing to the dripping regime occurs when the capillary number of the continuous phase, $\mathrm{Ca_c} \approx \eta_{\rm c} Q_{\rm c}/(\gamma h w)$ with $Q_{\rm c} \approx P_{\rm c}/(R_{\rm c} + R_{\rm x})$, exceeds roughly $10^{-2}$~\cite{zhu2017passive}. For $50 \lesssim P_{\rm c} \lesssim 350$~\si{\milli\bar}, we have $0.01 \lesssim \mathrm{Ca_c} \lesssim 0.2$. Our experiments thus cover the dripping and squeezing regimes. 

\begin{figure*}[t!]
\centering
\includegraphics[scale=1]{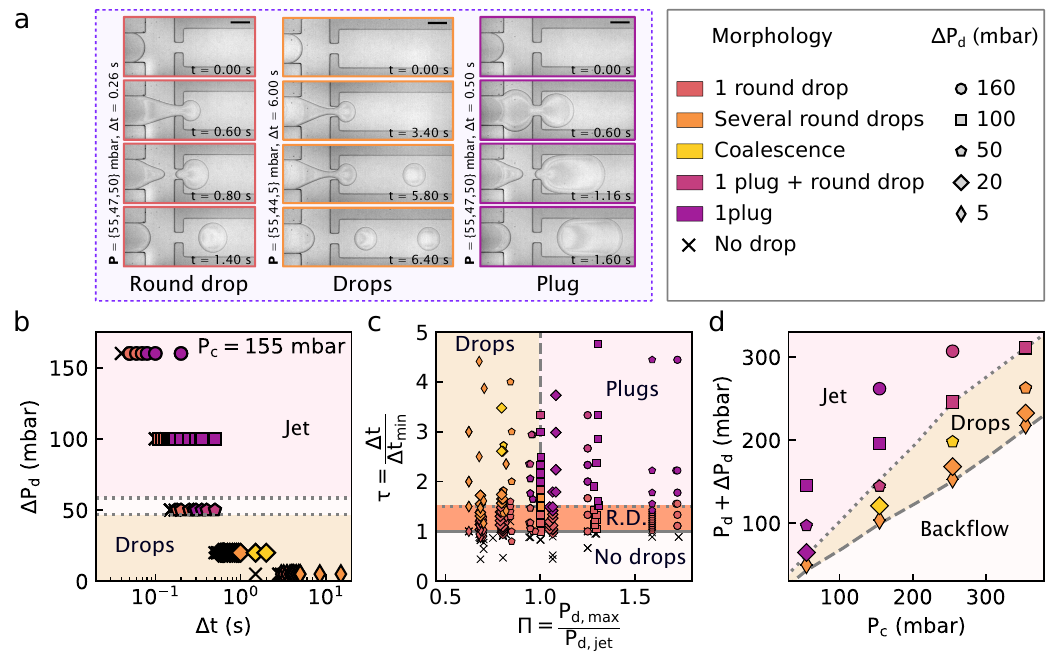}  
\caption{\textbf{Droplet Morphologies}. (a) Main morphologies observed for pulses with control parameters listed on the left of each image sequence; time is indicated inside each panel; scale bars: $100$~\si{\micro\meter}. %
(b-d) Diagrams showing DoD morphologies as they depend on various control parameters. Each graph shares the same legend (top-right); markers and colors respectively set the amplitude $\Delta P_\mathrm{d}$ and  the morphology. %
(b) Influence of the amplitude and duration of the pulse for $\{P_\mathrm{c}, P_\mathrm{d}\} = \{155, 100\}$~\si{\milli\bar}. The gap between the dotted lines indicates the experimental uncertainty on $P_{\rm d,jet} = 152 \pm 5$~\si{\milli\bar}. (c) Phase diagram as a function of dimensionless duration $\tau$ and pressure $\Pi$. (d) Phase diagram of droplet morphology as a function of the input pressures. Each datum corresponds to a set of experiments with fixed $\mathbf{P}$ and its color matches the morphology observed when $\Delta t$ is much larger than the minimal droplet production time; the y-coordinate corresponds to the pressure during the pulse, \emph{i.e.}, $P_\mathrm{d} + \Delta P_\mathrm{d}$. Dashed and dotted lines in (c) and (d) serve as guides to the eye to separate droplet and jet regimes from the steady conditions described in Section~\ref{SectSteady}. }
\label{fig:Morpho}
\end{figure*}

Lastly, if $P_{\rm d}$ is above a second threshold $P_{\rm d,jet}$, a continuous-phase jet  ---whose stability is owing to the out-of-plane confinement~\cite{guillot2008stability}--- flows towards the outlet. In Fig.~\ref{fig:setup}(b), the grey dotted line and the pink patch respectively denote this transition and the jetting regime. To retrieve the threshold $P_{\rm d,jet}$ theoretically, we follow Teo \emph{et al.}~\cite{teo2017negative} who described the biphasic flow of a jet using the hydraulic/electrical analogy. These authors approximate the hydraulic resistance of the dispersed-phase jet inside the exit channel $R_{\rm d,x}$ as that of a rectangular channel with the same width. Likewise, $R_{\rm c,x}$ stands for the hydraulic resistance of the continuous phase comprised between the dispersed phase and the wall of the chip. Hence, they separately apply the Hagen-Poiseuille law to each phase, which gives $P_\mathrm{c} = (R_{\rm c} + R_{\rm c,x}/2) \, Q_\mathrm{c} $ and $P_\mathrm{d} = (R_{\rm d} + R_{\rm d,x}) \, Q_\mathrm{d} $. Considering that the pressure across the section is invariant in the cross-stream direction for a parallel flow,  we impose $Q_\mathrm{c} R_{\rm c,x}/2 = Q_\mathrm{d} R_{\rm d,x}$ and have 
\begin{equation}
P_{\rm d,jet} = \frac{P_\mathrm{c}}{2} \left( \dfrac{\dfrac{R_{\rm d}}{R_{\rm d,x}} + 1}{\dfrac{R_{\rm c}}{\rm R_{\rm c,x}} + \dfrac{1}{2}} \right)\ .
\label{eq:StabJet}
\end{equation} 
For each $P_{\rm c}$, we set the pressure in the dispersed phase at the onset of the jetting regime. We measure the widths of the water and oil jets $w_{\rm d}$ and $w_{\rm c}$ and estimate $R_{\rm d,x}$ and $R_{\rm c,x}$ using Eq. \ref{eq:HydroRes}. We then calculate $P_{\rm d,jet}$ thanks to Eq. \ref{eq:StabJet}; the latter agrees well with the experimental thresholds above which we observe a stable jet. After characterizing the different regimes in a steady pressure conditions, we now investigate the case of time-dependent pressure pulses.

\subsection{Morphologies of on-demand droplets \label{sec:morphology}}

In Fig.~\ref{fig:setup}(c) is shown the result of a pressure pulse with $\mathbf{P} = \{155,98,100\}$~\si{\milli\bar} and $\Delta t = 0.10$~\si{\second}: a single circular droplet, as seen from above, is formed. Furthermore, shown in Fig.~\ref{fig:Morpho}(a) are three typical instances for which droplets are created upon the application of pressure pulses with $\mathbf{P} = \{55,47,50\}, \{55,44,5\}, \{55,47,50\}$~\si{\milli\bar} and $\Delta t = 0.26, 6.00, 0.50$~\si{\second}. Respectively, the result of these pulses are: one circular droplet (first column) as in Fig.~\ref{fig:setup}(c), but with a different pressure set; two circular droplets (second column); and finally, one long plug (third column). In this section, we present an overview of these different droplet morphologies. Then, we discuss how such observed morphologies evolve as a function of the pulse parameters, ultimately being controlled by where the pulse ends with respect to the steady-flow phase diagram shown in Fig.~\ref{fig:setup}(b), and the pulse duration. 

Once the stability of the meniscus is established with a choice of $\{P_{\rm c},P_{\rm d}\}$, droplets may be produced either by controlling the duration of the pressure pulse $\Delta t$, or by pushing with a greater amplitude $\Delta P_\mathrm{d}$. Concerning the pulse duration, for any pressure triplet $\mathbf{P}$, there exists a minimal pulse duration $\Delta t_{\rm min}$ below which no droplet is created. In this $\tau = \Delta t/\Delta t_{\rm min}<1$ regime, the dispersed phase fills a portion of the junction, but returns to the equilibrium position; such events are indicated with black crosses in Fig.~\ref{fig:Morpho}(b) for the specific case of $\{P_\mathrm{c}, P_\mathrm{d}\} = \{155, 100\}$~\si{\milli\bar} and several couples of $\{\Delta P_\mathrm{d}, \Delta t\}$. When $\Delta t\gtrsim\Delta t_{\rm min}$ is just reached at fixed $\mathbf{P}$, a single, round and pancake-shaped droplet stems from the pulse as shown in the first panel on the left in Fig.~\ref{fig:Morpho}(a). By contrast, as shown in the following panels, longer pulses may either produce several round, pancake-shaped droplets, or one elongated droplet that we refer to as a plug.

Indeed, as shown in the pressure-time phase diagram of Fig.~\ref{fig:Morpho}(c), we observe four regions. The first two are described in the previous paragraph, demarcating no droplets for $\tau \leq 1$ and single droplets, with pulses limited to the range $1 \leq \tau \lesssim 1.5$. The third region exists for what we call long pulses, roughly when $\tau \gtrsim 1.5$. Such pulses last long enough to generate several droplets or an elongated plug. Referring to the steady-phase diagram in Fig.~\ref{fig:setup}(b), we define the reduced pressure $\Pi = P_\mathrm{d,max}/P_\mathrm{d,jet}$ where $P_\mathrm{d,max}=P_\mathrm{d} + \Delta P_\mathrm{d}$. We note that for $\Pi<1$ several droplets are produced for long pulses, while for $\Pi>1$, plugs may be formed.

In order to understand these long-pulse morphologies, Fig.~\ref{fig:Morpho}(d) displays morphological indications for long pulses, superposed on the steady-condition phase diagram; each datum has coordinates $\{P_{\rm c}, P_\mathrm{d,max}\}$, with the color corresponding to the observed morphology (legend at top right). Long pulses produce plugs when the maximal pressure during the pulse falls in the jet regime, \emph{i.e.}, when $P_\mathrm{d,max} > P_{\rm d,jet}$. Conversely, when $P_\mathrm{d,max} < P_{\rm d,jet}$, the mechanism responsible for droplet production in steady conditions~\cite{zhu2017passive} limits the size of the droplet: in this pressure regime, long pulses result in the creation of multiple droplets. We thus generally observe a transposition of the steady phase diagram of coordinates $\{P_\mathrm{c}, P_{\rm d}\}$ into the DoD phase diagram of coordinates $\{P_\mathrm{c},P_\mathrm{d,max}\}$. 

While the previous discussion focused on the main cases of single droplet, plug and double droplet production, we note that the legend in Fig.~\ref{fig:Morpho} also contains some marginal cases. In particular, long pulses with $\tau>1.5$ may generate several droplets that eventually merge together. This merging is due to the relaxation of the upstream droplet as it exits the confinement of the channel while it is in contact with the downstream one. The pressure imbalance between the front and back of the droplet in the continuous phase results in drainage of the fluid in between the droplets, promoting coalescence~\cite{Liu2007, Tan2007, Babahosseini2018}. Also, when $\mathbf{P}$ falls into the jet regime, long pulses may produce additional round droplets when $P_{\rm c}$ is at the larger end of our investigated range. These two cases are respectively denoted as ``coalescence'' and ``plug $+$ round drop'' in Fig.~\ref{fig:Morpho}. Despite these marginal cases, once a pressure set and pulse time are chosen, the resulting morphologies are reproducible. To demonstrate this reproducibility, in Fig.~\ref{fig:SizeDist} are shown the results of 100 iterations of four separate $\mathbf{P}, \Delta t$ couples. For each, all 100 iterations produced the morphologies indicated in Fig.~\ref{fig:SizeDist}(a), and with a reasonable monodispersity in projected area, $\mathcal{A}$, as shown in Fig.~\ref{fig:SizeDist}(b). In the following section we compare these results to those obtainable in steady conditions. 

\subsection{Comparison to steady droplet production}

In steady conditions, the applied pressures set both the droplet size and the throughput. For a given $P_{\rm c}$, both quantities increase as a function of $P_{\rm d}$ until the jetting pressure is reached~\cite{guillot2008stability, teo2017negative}; the size of the droplet is thus limited to that obtained just before the transition. In our view, the advantage of using pressure pulses could be threefold, depending on the end-user's application. 

\begin{figure}[t!]
\centering
\includegraphics[scale=1]{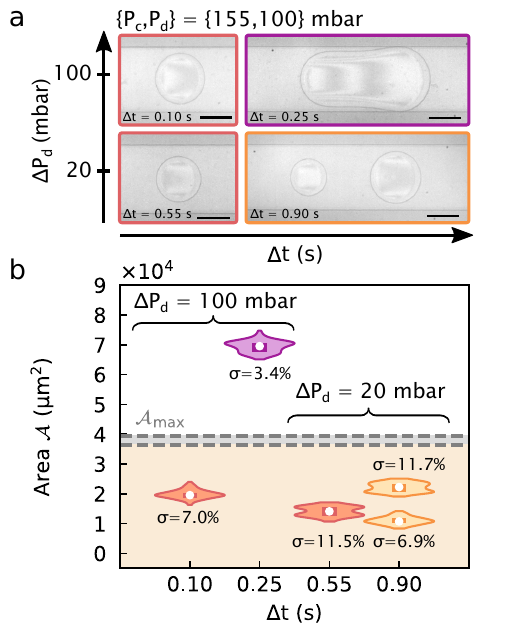}  
\caption{\textbf{Reproducibility of experiments}. (a) Top views obtained at two different pressure pulse amplitudes, starting from the same initial conditions $\{P_{\rm c},P_{\rm d}\}$. Each picture displays the resulting droplet for distinct values of $\Delta t$, as indicated. Short pulses generate a single round drop. Longer pulses either generate two round droplets or one plug. Plugs are obtained at high amplitude ($\Delta P_\mathrm{d} = 100$~\si{\milli\bar}). Scale bars: $100$~\si{\micro\meter}. (b) Corresponding area distributions and the standard deviation $\sigma$ when the experiments are iterated $100$ times. Each distribution displays its median (white dot) and its first and third quantiles (dark patch). Steady conditions upper area $\mathcal{A}_{\rm max} = 3.78 \pm 0.15 \times 10^4$~\si{\square\micro\meter} when $P_{\rm c} = 155$~\si{\milli\bar} (grey patch). 
}
\label{fig:SizeDist}
\end{figure}

First, in a pulsed generation, the volume injected and the production rate are controlled independently. In Fig.~\ref{fig:SizeDist} are shown the area distributions (viewed from the top) of single round droplets, plugs and two round droplets for two different pressure jumps and for different pulse durations. In each case, 100 independent pulses were used to produce droplets at intervals of several 10's of seconds to minutes. By increasing the waiting time between two pulses and choosing suitable $\mathbf{P}$ and $\Delta t$ so that only one droplet results from the pulse, the production rate can be tuned as low as desired. This on-demand production is particularly suitable for applications that require low generation frequency or sporadic droplet production. Low frequencies could be desired for example when it is required to have an exit channel that contains a single droplet only (\emph{i.e.} controlling the exit-channel resistance~\cite{Vanapalli2009}). Sporadic production may be required when the production of a droplet should be triggered by a rare event occurring in the continuous phase, this latter being monitored \emph{in situ}. Moreover, the size of the droplet can be tuned by keeping the same pressure conditions $\mathbf{P}$ (\emph{cf.} the two distributions at $\Delta P_{\rm d} = 100$~\si{\milli\bar}), but changing the pulse duration.

Second, using pressure pulses makes it possible to create on-demand droplets that are inaccessible under steady conditions. The orange patch in Fig.~\ref{fig:SizeDist}(b) represents the accessible sizes available in steady conditions. If the pulse amplitude is set too low ($\Delta P_{\rm d} = 20$~\si{\milli\bar}), $\mathbf{P}$ falls into the droplet production regime. The droplet sizes are still below the maximum projected droplet area in steady conditions, called $\mathcal{A}_{\rm max}$ (grey patch in Fig.~\ref{fig:SizeDist}(b)). Nevertheless, we can obtain droplets on-demand when we increase the amplitude up to $100$~\si{\milli\bar} for which similar steady conditions would give a stable jet. By increasing $\Delta t$, we can thus create individual plugs with projected areas larger than $\mathcal{A}_{\rm max}$.

While suitable for the creation of plugs not normally accessible under steady conditions, the DoD method presented here has some minor practical limitations.  While the morphology of the droplets are reproducible over $100$ iterations as described above, the size of the droplet is sensitive to the experimental conditions which may drift over long times. The relative error of each distribution is $\ 4\lesssim \sigma \lesssim 12$ \% which can be compared to $0.3\lesssim\sigma \lesssim 5$ \% as reported elsewhere~\cite{zhu2017passive, xu2006preparation, umbanhowar2000monodisperse, takeuchi2005axisymmetric, bransky2009microfluidic, hamidovic2020off, zhou2014facile, barkley2015}, for shorter time scales. Besides this drift, our pressure controller's response time of roughly $100~\si{\milli\second}$ gives a maximal frequency of order $10\ \si{\hertz}$~\cite{hamidovic2020off}. Other time-dependent strategies exceeding this limit may be possible, including, for example, piezoelectric actuation. Furthermore, and while not reported here, we have also implemented \emph{in-situ}, optical meniscus monitoring which can be used in PID control to stabilise the meniscus, reducing variability due to pressure drifts and the resulting drifts in the initial meniscus position; in practice and for the chips used here, we noted typical variations of a few mbar on~$P_\mathrm{d}$. 

Lastly, since DoD is principally reliant on a limitation of the supply of dispersed phase, we hypothesize that the method could be applied to higher viscosity ratios~\cite{Cubaud2008} with appropriate modifications to the pulse width and duration. In this regard, the DoD method could be compared to continuous-generation methods relying on an instability in the meniscus of the dispersed phase~\cite{Song2006} or phase inversion~\cite{Li2018}.

\subsection{Scaling of the droplet volume and minimal generation time}
Having described the qualitative aspects of our DoD strategy in the previous sections, in this section, we attempt to rationalize two aspects of the droplets created during a pressure pulse: in the first case, the droplet volumes, and in the second case, a simple scaling argument for the minimal pulse time. Our arguments are based on the electronic-hydraulic circuit analogy common to many microfluidic applications~\cite{BruusTheoreticalMicrofluidics2007}, which will allow us to predict the flow rate of the dispersed phase after a pulse is applied. 

\begin{figure*}[t]
\centering
\includegraphics[scale=1]{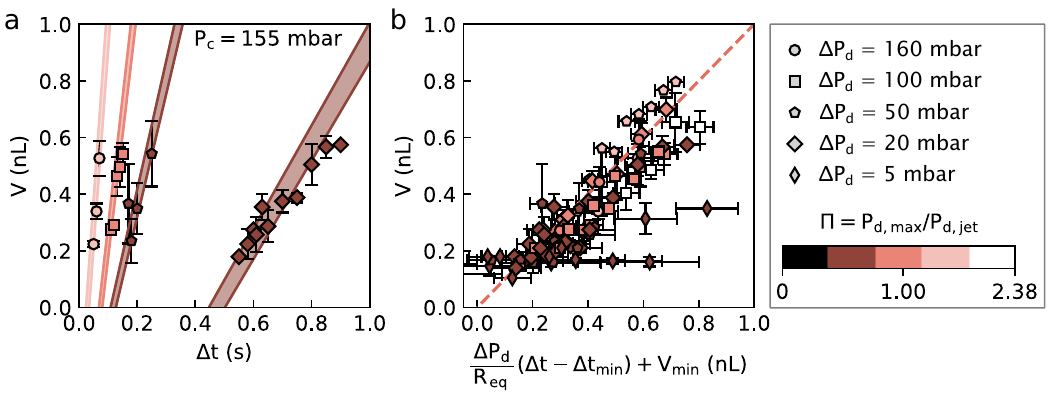}  
\caption{\textbf{Droplet volume scaling.} Volume of single round droplets. (a) Increase in volume $V$ as the pulse lasts longer, for different amplitudes. Here, $P_{\rm c} = 155$~\si{\milli\bar}. The color patches correspond to Eq. \ref{eq:VolScal} and take into account errors in the hydraulic resistances and the measurements of $\Delta t_{\rm min}$ and $V_{\rm min}$. The color refers to the non-dimensionnal pressure $\Pi$.
(b) Comparison of each $\mathbf{P} = \{P_{\rm c}$,$P_{\rm d}$,$\Delta P_{\rm d}\}$ dataset with the scaling in Eq.~\ref{eq:VolScal}. The dashed line accounts for the slope $1$. The scaling fails to capture the data only when $\Pi < 1$ (\textit{i.e.}, when $\mathbf{P}$ falls into the droplet production regime for long pulses).}
\label{fig:VolScal}
\end{figure*}
To predict the dispersed-phase flow rate, $Q_{\rm d}$, we refer to Fig.~\ref{fig:CrossSketch} where a typical flow-focusing geometry is shown at the instant just after the pulse has been initiated. When the pulse is on, the pressure drop across the dispersed phase writes $(P_{\rm d,max}- P_{\rm j^{-}})=R_{\rm d}Q_{\rm d}$. Likewise, the pressure drop of the continuous-phase  entrance channel writes $P_{\rm c} - P_{\rm j^{+}} = R_{\rm c}Q_{\rm c}$. Here $P_{\rm j^{-}}$ (resp. $P_{\rm j^{+}}$) stands for the pressure at the junction on the dispersed (resp. continuous) side. Writing the Laplace pressure across the interface $P_{\rm j^{-}} - P_{\rm j^{+}} = \Delta P_{\rm L}$ and equating the pressures at the junction gives: 
\begin{equation}
P_{\rm d,0} + \Delta P_{\rm d} - Q_{\rm d}R_{\rm d} - \Delta P_{\rm L} = P_{\rm c} - Q_{\rm c}R_{\rm c}\ .
\label{eq:DynMenCond}
\end{equation}
We consider in this scaling the early stage of the pulse; only the continuous phase flows inside the exit channel. Volume conservation imposes that the volume of dispersed phase injected in the junction pushes the same volume of continuous phase towards the outlet. Thus, the flow rate in the exit channel gives $P_{\rm j^{+}}/R_{\rm x} = Q_{\rm d} + Q_{\rm c}$. As $P_{\rm j^{+}} = P_{\rm c} - Q_{\rm c}R_{\rm c}$, we obtain $Q_{\rm c}$ as a function of $Q_{\rm d}$:
\begin{equation}
Q_{\rm c} = \dfrac{P_{\rm d} - Q_{\rm d} R_{\rm x}}{R_{\rm c} + R_{\rm x}}.  
\label{eq:QcQd}
\end{equation} 
Injecting Eq. \ref{eq:QcQd} into Eq. \ref{eq:DynMenCond}, and using the equilibrium conditions of the meniscus (Eq. \ref{eq:StabMen}), we obtain:
\begin{equation}
Q_{\rm d  } =   \dfrac{\Delta P_{\rm d}}{R_{\rm eq}}
\label{eq:Qd}
\end{equation} 
with:%
\begin{equation}
R_{\rm eq} = R_{\rm d} + \left(\dfrac{1}{R_{\rm c}}+\dfrac{1}{R_{\rm x}}\right)^{-1}.
\label{EqREQ}
\end{equation}
Using this circuit analysis, it will be possible to predict both the droplet volumes and the minimal droplet production time, as discussed next. We caution that this simple analysis neglects many details related to the meniscus friction~\cite{Cantat2013}, yet we have verified this latter to be negligible compared to the channel resistance friction, and the scaling analysis to follow captures the main features of the data presented below. Furthermore, the pulse amplitude $\Delta P_\mathrm{d}$ and duration $\Delta t$ required for drop generation and resulting volume reported below are specific to our chip geometry and working fluids. Nevertheless, the qualitative trends we report below should be recoverable in similar systems.

\subsubsection{Droplet volumes}
Referring to Fig.~\ref{fig:SizeDist}(a), the volume of a droplet in the exit channel can be defined approximately as $V= h \mathcal{A}$, a good approximation since the channel height is much less than the lateral dimensions of the droplets. Measurements of the volume were made for conditions under which a single round droplet stems from the pulse of the pressure. These volumes are shown as a function of the pulse duration in Fig.~\ref{fig:VolScal} for the particular case of $P_{\rm c} = 155$~\si{\milli\bar}, and several $\Delta P_{\rm d}$ as noted in the legend. For fixed initial pressures and amplitude, $V$ is a linear function of $\Delta t$ as observed in Fig.~\ref{fig:VolScal}(a). The slope, $\Delta V /\Delta t$, has the dimensions of a flow rate, which we hypothesise is well approximated by the flow rate of the dispersed phase predicted in Eq.~\ref{eq:QcQd}. As suggested by this latter equation, the flow rate increases with $\Delta P_{\rm d}$ for given initial conditions $\{P_{\rm c}$,$P_{\rm d}\}$, a result that is consistent with the data. 

Knowing that a $\Delta t_{\rm min}$-long pulse creates a droplet of volume $V_{\rm min}$, we integrate Eq.~\ref{eq:Qd} to propose the relationship:
\begin{equation}
V =   \frac{\Delta P_{\rm d}}{R_{\rm eq}} (\Delta t - \Delta t_{\rm min}) + V_{\rm min}.
\label{eq:VolScal}
\end{equation} 
In this equation, the resistances necessary for the prediction of $R_{\rm eq}$ according to Eq.~\ref{EqREQ} are shown in Table~\ref{tab:ChipGeo} while $\Delta t_\mathrm{min}$ and $V_\mathrm{min}$ can be measured experimentally. 

In Fig.~\ref{fig:VolScal}(b) are shown droplet volumes as a function of the scaled and shifted time for all of our data over a wide range of $\mathbf{P}$. Fig.~\ref{fig:VolScal}(a), with all data sets showing linearity, and (b) collapsing all of the data, show that Eq.~\ref{eq:VolScal} captures all of the $V = f(\mathbf{P},\Delta t)$ data, having measured the offsets experimentally. We find a particularly good agreement when $\mathbf{P}$ falls into the jet regime in steady conditions ({\it i.e.}, $\Pi >1$) shown with light red data sets in Fig.~\ref{fig:VolScal}(b), for which the departure from $\Delta t_\mathrm{min}$ is the largest. For situations of smaller $\Delta t$, our model neglects any generation time shorter than the pulse duration: the dispersed phase flows towards the outlet while the pulse is on, and the pinch-off occurs after the pulse ends. Consistently, the scaling does not hold at low amplitudes of $\Delta P_{\rm d}$, that is, when $\mathbf{P}$ falls close to the stable meniscus conditions ({\it i.e.}, $\Pi < 1$, darker datasets in Fig.~\ref{fig:VolScal}(b)). In this case, the mechanism responsible for pinch-off in steady conditions limits the size of the droplet.

\subsubsection{Minimal duration of the pulse}
\label{sec:DtMin}
\begin{figure*}[t!]
\centering
\includegraphics[scale=1]{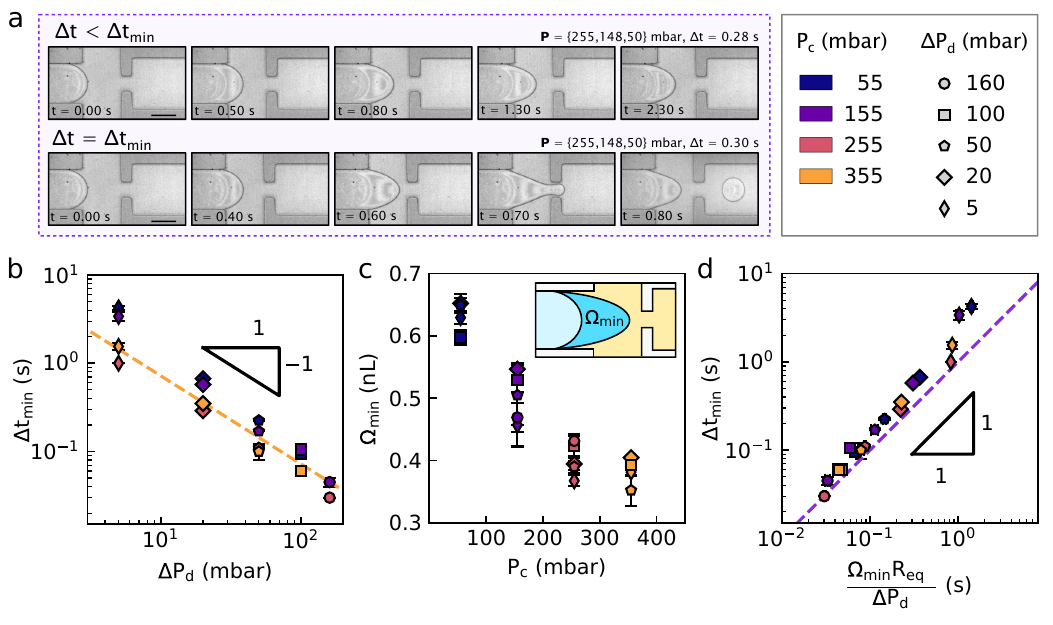}  
\caption{\textbf{Shortest pulse that generates a drop.} (a)  Pressure pulses at $P_{\rm c} = 255$~\si{\milli\bar} and $\Delta P_{\rm d} = 50$~\si{\milli\bar}, with $\Delta t<\Delta t_\mathrm{min}$ (top) and $\Delta t\simeq\Delta t_\mathrm{min}$ (bottom). Each picture displays the elapsed time in seconds. Scale bars: $100$~\si{\micro\meter}. (b) Minimal droplet generation time, $\Delta t_{\rm min}$, as a function of the pressure pulse amplitude, $\Delta P_{\rm d}$; the dashed line shows the prediction of Eq.~\ref{eq:DtMinScaleFirst}.  (c) The volume $\Omega_{\rm min}$, schematically indicated in the inset, injected in the junction for the longest pulse that does not generate a droplet, as a function of $P_{\rm c}$. (d) Minimal pulse duration as a function of the prediction of Eq.~\ref{eq:DtMinScaleFirst}, with $\Omega_{\rm j}$ replaced by the measured $\Omega_{\rm min}$. The dashed line represents the function $y = x$. }
\label{fig:DtMin}
\end{figure*}

In the previous section, we noted that creating a droplet requires a minimal $\Delta t_{\rm min}$ for any $\mathbf{P} = \{P_{\rm c}$, $P_{\rm d}, \Delta P_{\rm d}\}$. Indeed, in Fig.~\ref{fig:DtMin}(a) is shown the situation for $\Delta t < \Delta t_{\rm min}$ (top) and for $\Delta t \approx \Delta t_{\rm min}$ (bottom). For this former case, the dispersed phase flows towards the junction during the pulse, but the interface recedes back after the dispersed phase pressure returns to $P_{\rm d}$. In this section, the influence of $\mathbf{P}$ on $\Delta t_{\rm min}$ is described. 

As shown in Fig.~\ref{fig:DtMin}(b), the minimal pulse duration to create a droplet, $\Delta t_{\rm min}$, is inversely proportional to $\Delta P_{\rm d}$ for fixed initial conditions $\{P_{\rm c}$, $P_{\rm d}\}$. The flow rate of the dispersed phase increases with pulse amplitude, as in Eq.~\ref{eq:Qd}, so it takes less time for the dispersed phase to enter the continuous flow toward the outlet. Fig.~\ref{fig:DtMin}(b) also shows that $\Delta t_{\rm min}$ decreases with $P_{\rm c}$, although the influence of the latter is weaker than that of $\Delta P_{\rm d}$.
 
To rationalize the inverse relationship of Fig.~\ref{fig:DtMin}(b), we observe that for pulses slightly shorter than $\Delta t_{\rm min}$, a tongue of water fills most of the junction of volume $\Omega_{\rm j} = w^2h\approx 0.6$~nL before receding. This observation provides a flow rate of $\Omega_{\rm j}/\Delta t_{\rm min}$. According to Eq.~\ref{eq:Qd}, $Q_{\rm d} = \Delta P_{\rm d}/R_{\rm eq}$, thus equating the two rates we obtain:
\begin{equation}
\Delta t_{\rm min} \approx \frac{\Omega_{\rm j} R_{\rm eq}}{ \Delta P_{\rm d}}\ .
\label{eq:DtMinScaleFirst}
\end{equation} 
This scaling captures well the trend $\Delta t_{\rm min} \propto \Delta P_{\rm d}^{-1}$ as shown with the dashed line in Fig. \ref{fig:DtMin}(b). However, the weaker dependance on $P_{\rm c}$ remains uncaptured. 

We define $\Omega_{\rm min}$ as the volume difference between the initial position of the interface and its position just before it recedes for the minimal time, as shown schematically in the inset of Fig.~\ref{fig:DtMin}(c). In Fig.~\ref{fig:DtMin}(c), it is shown that the measured values of $\Omega_{\rm min}$ decrease with increasing $P_{\rm c}$, and that all $\Omega_{\rm min}$ are of order $\Omega_{\rm j}$. Indeed, increasing $P_{\rm c}$ at constant  $\Delta P_{\rm d}$ increases the flow rate ratio $Q_{\rm c}/Q_{\rm d}$; therefore, the continuous phase advances less into the junction before a droplet is formed.  

Replacing $\Omega_{\rm j}$ in Eq.~\ref{eq:DtMinScaleFirst} by the measured values of $\Omega_{\rm min}$ in Fig.~\ref{fig:DtMin}(d), we find an improved agreement with the scaling. We conclude that if the pulse duration is long enough such that Eq.~\ref{eq:Qd} predicts an integrated volume close to that of the junction, $\Omega_\mathrm{j}$, a droplet may be created. 

\section*{Conclusion}
In this paper, we applied a positive pressure pulse to the dispersed-phase inlet of a flow-focusing chip to produce droplets on-demand. We observed the morphology and size of the resulting droplets upon variations in pulse amplitude, pulse duration and initial pressures. The main result of our study is that droplets of controlled-size and generation period can be created in a wide range of pressure conditions, particularly here in the dripping and squeezing regimes. The droplet morphologies and topologies are well-described in the context of the phase diagram in steady conditions, and to the typical droplet generation time in the appropriate space of steady dispersed and continuous pressures.  When the  pressure during the pulse falls inside the coflowing-jet regime, it is possible to generate elongated droplets that overcome the volume limitation in steady conditions. Using the electrical-hydraulic analogy, we predicted the volume of the droplets as well as the minimal duration to create a droplet as a function of the hydraulic resistances of the chip and the amplitude of the pulse. Our simple theoretical scaling argument does not include the influence of the continuous pressure on minimal tongue volume and could be refined by considering the details of the meniscus hydrodynamics in the junction. Nevertheless, we provide updated insights for the generation of droplets on-demand by means of a cheap and easy-to-implement method. The droplet volumes and their generation time being determined by the chip geometry and pulse duration. This method can: be adapted to chips with several inlets and used, for example, to synchronise object production inside a microfluidic network; to decouple droplet size from generation frequency; or, to create controlled polydispersity (\emph{e.g.}, bi- or tri-disperse) in droplet ensembles. 

\acknowledgements{The authors acknowledge the financial support of the Institut Pierre-Gilles de Gennes (Equipex ANR-10-EQPX-0034 and Labex ANR-10-LABX-31), the PSL-Qlife Prematuration and PSL Maturation grants (C20-70/2020-105, C22-89-2022-429, through the Investissement d'Avenir ANR-17-SATE-0002), the BFT-lab program (Banque Publique d'investissement). The authors were also partially supported by the “Springboard programme for a sustainable future” funded by the British Council in France. The work lastly benefited from the technical contribution of the joint service unit CNRS UAR 3750; the authors are very grateful for the continued support of the engineers on this unit.}

\vspace{1mm}

\noindent \textbf{Author contribution statement} MO performed and designed experiments, wrote image analysis codes, developed the modelling, wrote the first version of and edited the manuscript. GC, SH-C, MG and FB performed and designed experiments,  and edited the manuscript. ML and JDM designed experiments, developed the modelling and edited the manuscript. 

\vspace{1mm}

\noindent \textbf{Data availability} The datasets generated during the current study are available from the corresponding authors upon reasonable request.


\begin{thebibliography}{0}%
\makeatletter
\providecommand \@ifxundefined [1]{%
 \@ifx{#1\undefined}
}%
\providecommand \@ifnum [1]{%
 \ifnum #1\expandafter \@firstoftwo
 \else \expandafter \@secondoftwo
 \fi
}%
\providecommand \@ifx [1]{%
 \ifx #1\expandafter \@firstoftwo
 \else \expandafter \@secondoftwo
 \fi
}%
\providecommand \natexlab [1]{#1}%
\providecommand \enquote  [1]{``#1''}%
\providecommand \bibnamefont  [1]{#1}%
\providecommand \bibfnamefont [1]{#1}%
\providecommand \citenamefont [1]{#1}%
\providecommand \href@noop [0]{\@secondoftwo}%
\providecommand \href [0]{\begingroup \@sanitize@url \@href}%
\providecommand \@href[1]{\@@startlink{#1}\@@href}%
\providecommand \@@href[1]{\endgroup#1\@@endlink}%
\providecommand \@sanitize@url [0]{\catcode `\\12\catcode `\$12\catcode
  `\&12\catcode `\#12\catcode `\^12\catcode `\_12\catcode `\%12\relax}%
\providecommand \@@startlink[1]{}%
\providecommand \@@endlink[0]{}%
\providecommand \url  [0]{\begingroup\@sanitize@url \@url }%
\providecommand \@url [1]{\endgroup\@href {#1}{\urlprefix }}%
\providecommand \urlprefix  [0]{URL }%
\providecommand \Eprint [0]{\href }%
\providecommand \doibase [0]{https://doi.org/}%
\providecommand \selectlanguage [0]{\@gobble}%
\providecommand \bibinfo  [0]{\@secondoftwo}%
\providecommand \bibfield  [0]{\@secondoftwo}%
\providecommand \translation [1]{[#1]}%
\providecommand \BibitemOpen [0]{}%
\providecommand \bibitemStop [0]{}%
\providecommand \bibitemNoStop [0]{.\EOS\space}%
\providecommand \EOS [0]{\spacefactor3000\relax}%
\providecommand \BibitemShut  [1]{\csname bibitem#1\endcsname}%
\let\auto@bib@innerbib\@empty
\end{thebibliography}%


\begin{thebibliography}{10}

\bibitem{seemann2011droplet}
R.~Seemann, M.~Brinkmann, T.~Pfohl, and S.~Herminghaus, ``Droplet based
  microfluidics,'' {\em Reports on progress in physics}, vol.~75, no.~1,
  p.~016601, 2011.

\bibitem{zhu2017passive}
P.~Zhu and L.~Wang, ``Passive and active droplet generation with microfluidics:
  a review,'' {\em Lab on a Chip}, vol.~17, no.~1, pp.~34--75, 2017.

\bibitem{datta201425th}
S.~S. Datta, A.~Abbaspourrad, E.~Amstad, J.~Fan, S.-H. Kim, M.~Romanowsky,
  H.~C. Shum, B.~Sun, A.~S. Utada, M.~Windbergs, {\em et~al.}, ``25th
  anniversary article: Double emulsion templated solid microcapsules: Mechanics
  and controlled release,'' {\em Advanced Materials}, vol.~26, no.~14,
  pp.~2205--2218, 2014.

\bibitem{lee2016microfluidic}
T.~Y. Lee, T.~M. Choi, T.~S. Shim, R.~A. Frijns, and S.-H. Kim, ``Microfluidic
  production of multiple emulsions and functional microcapsules,'' {\em Lab on
  a Chip}, vol.~16, no.~18, pp.~3415--3440, 2016.

\bibitem{nunes2013dripping}
J.~Nunes, S.~Tsai, J.~Wan, and H.~A. Stone, ``Dripping and jetting in
  microfluidic multiphase flows applied to particle and fibre synthesis,'' {\em
  Journal of physics D: Applied physics}, vol.~46, no.~11, p.~114002, 2013.

\bibitem{kong2012droplet}
T.~Kong, J.~Wu, M.~To, K.~Wai Kwok~Yeung, H.~Cheung~Shum, and L.~Wang,
  ``Droplet based microfluidic fabrication of designer microparticles for
  encapsulation applications,'' {\em Biomicrofluidics}, vol.~6, no.~3, 2012.

\bibitem{song2008microfluidic}
Y.~Song, J.~Hormes, and C.~S. Kumar, ``Microfluidic synthesis of
  nanomaterials,'' {\em small}, vol.~4, no.~6, pp.~698--711, 2008.

\bibitem{jahn2008preparation}
A.~Jahn, J.~E. Reiner, W.~N. Vreeland, D.~L. DeVoe, L.~E. Locascio, and
  M.~Gaitan, ``Preparation of nanoparticles by continuous-flow microfluidics,''
  {\em Journal of Nanoparticle Research}, vol.~10, pp.~925--934, 2008.

\bibitem{martz2012microfluidic}
T.~D. Martz, D.~Bardin, P.~S. Sheeran, A.~P. Lee, and P.~A. Dayton,
  ``Microfluidic generation of acoustically active nanodroplets,'' {\em Small},
  vol.~8, no.~12, pp.~1876--1879, 2012.

\bibitem{schroen2021droplet}
K.~Schroen, C.~Berton-Carabin, D.~Renard, M.~Marquis, A.~Boire, R.~Cochereau,
  C.~Amine, and S.~Marze, ``Droplet microfluidics for food and nutrition
  applications,'' {\em Micromachines}, vol.~12, no.~8, p.~863, 2021.

\bibitem{park2021microfluidic}
D.~Park, H.~Kim, and J.~W. Kim, ``Microfluidic production of monodisperse
  emulsions for cosmetics,'' {\em Biomicrofluidics}, vol.~15, no.~5, 2021.

\bibitem{theberge2010microdroplets}
A.~B. Theberge, F.~Courtois, Y.~Schaerli, M.~Fischlechner, C.~Abell,
  F.~Hollfelder, and W.~T. Huck, ``Microdroplets in microfluidics: an evolving
  platform for discoveries in chemistry and biology,'' {\em Angewandte Chemie
  International Edition}, vol.~49, no.~34, pp.~5846--5868, 2010.

\bibitem{vyawahare2010miniaturization}
S.~Vyawahare, A.~D. Griffiths, and C.~A. Merten, ``Miniaturization and
  parallelization of biological and chemical assays in microfluidic devices,''
  {\em Chemistry \& biology}, vol.~17, no.~10, pp.~1052--1065, 2010.

\bibitem{BruusTheoreticalMicrofluidics2007}
H.~Bruus, {\em Theoretical {{Microfluidics}}}.
\newblock Oxford {{Master Series}} in {{Physics}}, Oxford, New York: Oxford
  University Press, Sept. 2007.

\bibitem{guillot2008stability}
P.~Guillot, A.~Colin, and A.~Ajdari, ``Stability of a jet in confined
  pressure-driven biphasic flows at low reynolds number in various
  geometries,'' {\em Physical Review E}, vol.~78, no.~1, p.~016307, 2008.

\bibitem{garstecki2004formation}
P.~Garstecki, I.~Gitlin, W.~DiLuzio, G.~M. Whitesides, E.~Kumacheva, and H.~A.
  Stone, ``Formation of monodisperse bubbles in a microfluidic flow-focusing
  device,'' {\em Applied Physics Letters}, vol.~85, no.~13, pp.~2649--2651,
  2004.

\bibitem{takeuchi2005axisymmetric}
S.~Takeuchi, P.~Garstecki, D.~B. Weibel, and G.~M. Whitesides, ``An
  axisymmetric flow-focusing microfluidic device,'' {\em Advanced materials},
  vol.~17, no.~8, pp.~1067--1072, 2005.

\bibitem{yobas2006high}
L.~Yobas, S.~Martens, W.-L. Ong, and N.~Ranganathan, ``High-performance
  flow-focusing geometry for spontaneous generation of monodispersed
  droplets,'' {\em Lab on a Chip}, vol.~6, no.~8, pp.~1073--1079, 2006.

\bibitem{tirandazi2017liquid}
P.~Tirandazi and C.~H. Hidrovo, ``Liquid-in-gas droplet microfluidics;
  experimental characterization of droplet morphology, generation frequency,
  and monodispersity in a flow-focusing microfluidic device,'' {\em Journal of
  Micromechanics and Microengineering}, vol.~27, no.~7, p.~075020, 2017.

\bibitem{lashkaripour2019performance}
A.~Lashkaripour, C.~Rodriguez, L.~Ortiz, and D.~Densmore, ``Performance tuning
  of microfluidic flow-focusing droplet generators,'' {\em Lab on a Chip},
  vol.~19, no.~6, pp.~1041--1053, 2019.

\bibitem{Sun2011}
R.~Sun and T.~Cubaud, ``Dissolution of carbon dioxide bubbles and microfluidic
  multiphase flows,'' {\em Lab on a Chip}, vol.~11, pp.~2924--2928, Aug. 2011.

\bibitem{Cubaud2021}
T.~Cubaud, B.~Conry, X.~Hu, and T.~Dinh, ``Diffusive and capillary
  instabilities of viscous fluid threads in microchannels,'' {\em Physical
  Review Fluids}, vol.~6, p.~094202, Sept. 2021.

\bibitem{DeBlois2021}
C.~de~Blois, V.~Bertin, S.~Suda, M.~Ichikawa, M.~Reyssat, and O.~Dauchot,
  ``Swimming droplets in {{1D}} geometries: An active {{Bretherton}} problem,''
  {\em Soft Matter}, vol.~17, pp.~6646--6660, July 2021.

\bibitem{Reichert2019}
B.~Reichert, I.~Cantat, and M.-C. Jullien, ``Predicting droplet velocity in a
  {{Hele-Shaw}} cell,'' {\em Physical Review Fluids}, vol.~4, p.~113602, Nov.
  2019.

\bibitem{Vanapalli2009}
S.~A. Vanapalli, A.~G. Banpurkar, D.~van~den Ende, M.~H.~G. Duits, and
  F.~Mugele, ``Hydrodynamic resistance of single confined moving drops in
  rectangular microchannels,'' {\em Lab on a Chip}, vol.~9, pp.~982--990, Apr.
  2009.

\bibitem{he2005electro}
M.~He, J.~S. Kuo, and D.~T. Chiu, ``Electro-generation of single femtoliter-and
  picoliter-volume aqueous droplets in microfluidic systems,'' {\em Applied
  Physics Letters}, vol.~87, no.~3, 2005.

\bibitem{he2006effects}
M.~He, J.~S. Kuo, and D.~T. Chiu, ``Effects of ultrasmall orifices on the
  electrogeneration of femtoliter-volume aqueous droplets,'' {\em Langmuir},
  vol.~22, no.~14, pp.~6408--6413, 2006.

\bibitem{zhang2008manipulations}
M.~Zhang, J.~Wu, X.~Niu, W.~Wen, and P.~Sheng, ``Manipulations of microfluidic
  droplets using electrorheological carrier fluid,'' {\em Physical Review E},
  vol.~78, no.~6, p.~066305, 2008.

\bibitem{niu2009generation}
X.~Niu, M.~Zhang, J.~Wu, W.~Wen, and P.~Sheng, ``Generation and manipulation of
  “smart” droplets,'' {\em Soft Matter}, vol.~5, no.~3, pp.~576--581, 2009.

\bibitem{gu2011microfluidic}
H.~Gu, C.~U. Murade, M.~H. Duits, and F.~Mugele, ``A microfluidic platform for
  on-demand formation and merging of microdroplets using electric control,''
  {\em Biomicrofluidics}, vol.~5, no.~1, 2011.

\bibitem{kahkeshani2016drop}
S.~Kahkeshani and D.~Di~Carlo, ``Drop formation using ferrofluids driven
  magnetically in a step emulsification device,'' {\em Lab on a Chip}, vol.~16,
  no.~13, pp.~2474--2480, 2016.

\bibitem{Wu2012PulsedLaser}
T.-H. Wu, Y.~Chen, S.-Y. Park, J.~Hong, T.~Teslaa, J.~F. Zhong, D.~Di~Carlo,
  M.~A. Teitell, and P.-Y. Chiou, ``Pulsed laser triggered high speed
  microfluidic fluorescence activated cell sorter,'' {\em Lab on a Chip},
  vol.~12, pp.~1378--1383, 2012.

\bibitem{collins2013surface}
D.~J. Collins, T.~Alan, K.~Helmerson, and A.~Neild, ``Surface acoustic waves
  for on-demand production of picoliter droplets and particle encapsulation,''
  {\em Lab on a Chip}, vol.~13, no.~16, pp.~3225--3231, 2013.

\bibitem{brenker2016chip}
J.~C. Brenker, D.~J. Collins, H.~Van~Phan, T.~Alan, and A.~Neild, ``On-chip
  droplet production regimes using surface acoustic waves,'' {\em Lab on a
  Chip}, vol.~16, no.~9, pp.~1675--1683, 2016.

\bibitem{brenker2020demand}
J.~C. Brenker, C.~Devendran, A.~Neild, and T.~Alan, ``On-demand sample
  injection: combining acoustic actuation with a tear-drop shaped nozzle to
  generate droplets with precise spatial and temporal control,'' {\em Lab on a
  Chip}, vol.~20, no.~2, pp.~253--265, 2020.

\bibitem{xu2008drop}
J.~Xu and D.~Attinger, ``Drop on demand in a microfluidic chip,'' {\em Journal
  of Micromechanics and Microengineering}, vol.~18, no.~6, p.~065020, 2008.

\bibitem{bransky2009microfluidic}
A.~Bransky, N.~Korin, M.~Khoury, and S.~Levenberg, ``A microfluidic droplet
  generator based on a piezoelectric actuator,'' {\em Lab on a Chip}, vol.~9,
  no.~4, pp.~516--520, 2009.

\bibitem{shemesh2011coalescence}
J.~Shemesh, A.~Nir, A.~Bransky, and S.~Levenberg, ``Coalescence-assisted
  generation of single nanoliter droplets with predefined composition,'' {\em
  Lab on a Chip}, vol.~11, no.~19, pp.~3225--3230, 2011.

\bibitem{galas2009active}
J.-C. Galas, D.~Bartolo, and V.~Studer, ``Active connectors for microfluidic
  drops on demand,'' {\em New Journal of Physics}, vol.~11, no.~7, p.~075027,
  2009.

\bibitem{lin2008demand}
B.-C. Lin and Y.-C. Su, ``On-demand liquid-in-liquid droplet metering and
  fusion utilizing pneumatically actuated membrane valves,'' {\em Journal of
  Micromechanics and Microengineering}, vol.~18, no.~11, p.~115005, 2008.

\bibitem{lee2009predictive}
W.~S. Lee, S.~Jambovane, D.~Kim, and J.~W. Hong, ``Predictive model on micro
  droplet generation through mechanical cutting,'' {\em Microfluidics and
  nanofluidics}, vol.~7, pp.~431--438, 2009.

\bibitem{zeng2009microvalve}
S.~Zeng, B.~Li, J.~Qin, B.~Lin, {\em et~al.}, ``Microvalve-actuated precise
  control of individual droplets in microfluidic devices,'' {\em Lab on a
  Chip}, vol.~9, no.~10, pp.~1340--1343, 2009.

\bibitem{churski2010droplet}
K.~Churski, J.~Michalski, and P.~Garstecki, ``Droplet on demand system
  utilizing a computer controlled microvalve integrated into a stiff polymeric
  microfluidic device,'' {\em Lab on a Chip}, vol.~10, no.~4, pp.~512--518,
  2010.

\bibitem{guo2010valve}
F.~Guo, K.~Liu, X.-H. Ji, H.-J. Ding, M.~Zhang, Q.~Zeng, W.~Liu, S.-S. Guo, and
  X.-Z. Zhao, ``Valve-based microfluidic device for droplet on-demand operation
  and static assay,'' {\em Applied Physics Letters}, vol.~97, no.~23, 2010.

\bibitem{moon2015microfluidic}
B.-U. Moon, S.~G. Jones, D.~K. Hwang, and S.~S. Tsai, ``Microfluidic generation
  of aqueous two-phase system (atps) droplets by controlled pulsating inlet
  pressures,'' {\em Lab on a Chip}, vol.~15, no.~11, pp.~2437--2444, 2015.

\bibitem{zhou2014facile}
H.~Zhou and S.~Yao, ``A facile on-demand droplet microfluidic system for
  lab-on-a-chip applications,'' {\em microfluidics and nanofluidics}, vol.~16,
  pp.~667--675, 2014.

\bibitem{churski2010high}
K.~Churski, P.~Korczyk, and P.~Garstecki, ``High-throughput automated droplet
  microfluidic system for screening of reaction conditions,'' {\em Lab on a
  Chip}, vol.~10, no.~7, pp.~816--818, 2010.

\bibitem{dolega2012iterative}
M.~E. Dolega, S.~Jakiela, M.~Razew, A.~Rakszewska, O.~Cybulski, and
  P.~Garstecki, ``Iterative operations on microdroplets and continuous
  monitoring of processes within them; determination of solubility diagrams of
  proteins,'' {\em Lab on a Chip}, vol.~12, no.~20, pp.~4022--4025, 2012.

\bibitem{yu2015demand}
M.~Yu, Y.~Hou, H.~Zhou, and S.~Yao, ``An on-demand nanofluidic concentrator,''
  {\em Lab on a Chip}, vol.~15, no.~6, pp.~1524--1532, 2015.

\bibitem{jakiela2014generation}
S.~Jakiela, P.~R. Debski, B.~Dabrowski, and P.~Garstecki, ``Generation of
  nanoliter droplets on demand at hundred-hz frequencies,'' {\em
  Micromachines}, vol.~5, no.~4, pp.~1002--1011, 2014.

\bibitem{jung2010demand}
S.-Y. Jung, S.~T. Retterer, and C.~P. Collier, ``On-demand generation of
  monodisperse femtolitre droplets by shape-induced shear,'' {\em Lab on a
  Chip}, vol.~10, no.~20, pp.~2688--2694, 2010.

\bibitem{hamidovic2020off}
M.~Hamidovic, U.~Marta, H.~Bridle, D.~Hamidovic, G.~Fink, R.~Wille,
  A.~Springer, and W.~Haselmayr, ``Off-chip-controlled droplet-on-demand method
  for precise sample handling,'' {\em ACS omega}, vol.~5, no.~17,
  pp.~9684--9689, 2020.

\bibitem{teo2017negative}
A.~J. Teo, K.-H.~H. Li, N.-T. Nguyen, W.~Guo, N.~Heere, H.-D. Xi, C.-W. Tsao,
  W.~Li, and S.~H. Tan, ``Negative pressure induced droplet generation in a
  microfluidic flow-focusing device,'' {\em Analytical chemistry}, vol.~89,
  no.~8, pp.~4387--4391, 2017.

\bibitem{fink2020design}
G.~Fink, M.~Hamidovi{\'c}, A.~Springer, R.~Wille, and W.~Haselmayr, ``Design
  and realization of flexible droplet-based lab-on-a-chip devices,'' {\em e \&
  i Elektrotechnik und Informationstechnik}, 2020.

\bibitem{totlani2020scalable}
K.~Totlani, J.-W. Hurkmans, W.~M. Van~Gulik, M.~T. Kreutzer, and V.~Van~Steijn,
  ``Scalable microfluidic droplet on-demand generator for non-steady operation
  of droplet-based assays,'' {\em Lab on a Chip}, vol.~20, no.~8,
  pp.~1398--1409, 2020.

\bibitem{xia1998soft}
Y.~Xia and G.~M. Whitesides, ``Soft lithography,'' {\em Annual review of
  materials science}, vol.~28, no.~1, pp.~153--184, 1998.

\bibitem{pascual2019wettability}
M.~Pascual, M.~Kerdraon, Q.~Rezard, M.-C. Jullien, and L.~Champougny,
  ``Wettability patterning in microfluidic devices using thermally-enhanced
  hydrophobic recovery of pdms,'' {\em Soft Matter}, vol.~15, no.~45,
  pp.~9253--9260, 2019.

\bibitem{garstecki2005mechanism}
P.~Garstecki, H.~A. Stone, and G.~M. Whitesides, ``Mechanism for flow-rate
  controlled breakup in confined geometries: A route to monodisperse
  emulsions,'' {\em Physical review letters}, vol.~94, no.~16, p.~164501, 2005.

\bibitem{dollet2008role}
B.~Dollet, W.~van Hoeve, J.-P. Raven, P.~Marmottant, and M.~Versluis, ``Role of
  the channel geometry on the bubble pinch-off in flow-focusing devices,'' {\em
  Physical review letters}, vol.~100, no.~3, p.~034504, 2008.

\bibitem{daerr2016pendent_drop}
A.~Daerr and A.~Mogne, ``Pendent\_drop: an imagej plugin to measure the surface
  tension from an image of a pendent drop,'' {\em Journal of Open Research
  Software}, vol.~4, no.~1, pp.~e3--e3, 2016.

\bibitem{oh2012design}
K.~W. Oh, K.~Lee, B.~Ahn, and E.~P. Furlani, ``Design of pressure-driven
  microfluidic networks using electric circuit analogy,'' {\em Lab on a Chip},
  vol.~12, no.~3, pp.~515--545, 2012.

\bibitem{cornish1928flow}
R.~J. Cornish, ``Flow in a pipe of rectangular cross-section,'' {\em
  Proceedings of the Royal Society of London. Series A, Containing Papers of a
  Mathematical and Physical Character}, vol.~120, no.~786, pp.~691--700, 1928.

\bibitem{Agresti2010}
J.~J. Agresti, E.~Antipov, A.~R. Abate, K.~Ahn, A.~C. Rowat, J.-C. Baret,
  M.~Marquez, A.~M. Klibanov, A.~D. Griffiths, and D.~A. Weitz,
  ``Ultrahigh-throughput screening in drop-based microfluidics for directed
  evolution,'' {\em Proceedings of the National Academy of Sciences}, vol.~107,
  pp.~4004--4009, Mar. 2010.

\bibitem{Jeong2012}
W.-C. Jeong, J.-M. Lim, J.-H. Choi, J.-H. Kim, Y.-J. Lee, S.-H. Kim, G.~Lee,
  J.-D. Kim, G.-R. Yi, and S.-M. Yang, ``Controlled generation of submicron
  emulsion droplets via highly stable tip-streaming mode in microfluidic
  devices,'' {\em Lab on a Chip}, vol.~12, no.~8, p.~1446, 2012.

\bibitem{jousse2005compact}
F.~Jousse, G.~Lian, R.~Janes, and J.~Melrose, ``Compact model for multi-phase
  liquid--liquid flows in micro-fluidic devices,'' {\em Lab on a Chip}, vol.~5,
  no.~6, pp.~646--656, 2005.

\bibitem{fuerstman2007pressure}
M.~J. Fuerstman, A.~Lai, M.~E. Thurlow, S.~S. Shevkoplyas, H.~A. Stone, and
  G.~M. Whitesides, ``The pressure drop along rectangular microchannels
  containing bubbles,'' {\em Lab on a Chip}, vol.~7, no.~11, pp.~1479--1489,
  2007.

\bibitem{schindler2008droplet}
M.~Schindler and A.~Ajdari, ``Droplet traffic in microfluidic networks: A
  simple model for understanding and designing,'' {\em Physical Review
  Letters}, vol.~100, no.~4, p.~044501, 2008.

\bibitem{guyard2022elastohydrodynamic}
G.~Guyard, F.~Restagno, and J.~D. McGraw, ``Elastohydrodynamic relaxation of
  soft and deformable microchannels,'' {\em Physical Review Letters}, vol.~129,
  no.~20, p.~204501, 2022.

\bibitem{Liu2007}
K.~Liu, H.~Ding, Y.~Chen, and X.-Z. Zhao, ``Droplet-based synthetic method
  using microflow focusing and droplet fusion,'' {\em Microfluidics and
  Nanofluidics}, vol.~3, pp.~239--243, Apr. 2007.

\bibitem{Tan2007}
Y.-C. Tan, Y.~L. Ho, and A.~P. Lee, ``Droplet coalescence by geometrically
  mediated flow in microfluidic channels,'' {\em Microfluidics and
  Nanofluidics}, vol.~3, pp.~495--499, Aug. 2007.

\bibitem{Babahosseini2018}
H.~Babahosseini, T.~Misteli, and D.~L. DeVoe, ``Active or {{Passive On-Demand
  Droplet Merging}} in a {{Microfluidic Valve-Based Trap}},'' in {\em 2018 40th
  {{Annual International Conference}} of the {{IEEE Engineering}} in
  {{Medicine}} and {{Biology Society}} ({{EMBC}})}, (Honolulu, HI, USA),
  pp.~5350--5353, IEEE, July 2018.

\bibitem{xu2006preparation}
J.~Xu, S.~Li, J.~Tan, Y.~Wang, and G.~Luo, ``Preparation of highly monodisperse
  droplet in a t-junction microfluidic device,'' {\em AIChE journal}, vol.~52,
  no.~9, pp.~3005--3010, 2006.

\bibitem{umbanhowar2000monodisperse}
P.~Umbanhowar, V.~Prasad, and D.~A. Weitz, ``Monodisperse emulsion generation
  via drop break off in a coflowing stream,'' {\em Langmuir}, vol.~16, no.~2,
  pp.~347--351, 2000.

\bibitem{barkley2015}
S.~Barkley, E.~R. Weeks, and K.~{Dalnoki-Veress}, ``Snap-off production of
  monodisperse droplets,'' {\em The European Physical Journal E}, vol.~38,
  p.~138, Dec. 2015.

\bibitem{Cubaud2008}
T.~Cubaud and T.~G. Mason, ``Capillary threads and viscous droplets in square
  microchannels,'' {\em Physics of Fluids}, vol.~20, p.~053302, May 2008.

\bibitem{Song2006}
H.~Song, D.~L. Chen, and R.~F. Ismagilov, ``Reactions in droplets in
  microfluidic channels,'' {\em Angewandte Chemie International Edition},
  vol.~45, no.~44, pp.~7336--7356, 2006.

\bibitem{Li2018}
J.~Li, J.~Man, Z.~Li, and H.~Chen, ``Fabricating {{High-viscosity Droplets}}
  using {{Microfluidic Capillary Device}} with {{Phase-inversion Co-flow
  Structure}},'' {\em Journal of Visualized Experiments}, 2018.

\bibitem{Cantat2013}
I.~Cantat, ``Liquid meniscus friction on a wet plate: {{Bubbles}}, lamellae,
  and foams,'' {\em Physics of Fluids}, vol.~25, p.~031303, Mar. 2013.

\end{thebibliography}

\end{document}